# Onset of transient convection in a porous medium with an embedded low-permeability layer

Emmanuel E. Luther[1], Seyed M. Shariatipour[1], Ran Holtzman[1], and Michael C. Dallaston[2]

1) *Fluid and Complex Systems Research Centre, Coventry University, United Kingdom*
2) *School of Mathematical Sciences, Queensland University of Technology, Brisbane, Australia*

**Abstract**

Deep saline aquifers used for $CO_2$ sequestration are commonly made of sedimentary formations consisting of several layers of distinguishable permeability. In this work, the effect of a non-monotonic, vertically varying permeability profile on the onset of convective instability is studied theoretically using linear stability analyses. The onset time depends on the interaction between the permeability profile and the location of the concentration perturbation peak beyond which the concentration of $CO_2$ decays. A thin low-permeability layer can either accelerate or delay the onset time of the convective instability depending on the nature of the permeability variation – whether the permeability transition is smooth or layered, the Rayleigh number ($Ra$), and the location of the permeability change ($\hat{a}$) relative to the perturbation peak ($\hat{a}_{c^*}$), which scales as $\hat{a}_{c^*} \approx 14 Ra^{-1}$ for homogeneous systems. However, the low permeable layer has no effect on the onset time when it is near the lower boundary of a medium with sufficiently large $Ra$ ($\hat{a}_{c^*} \ll \hat{a}$). This nontrivial dependence highlights the implication of ignoring geological features of a small spatial extent, indicating the importance of a detailed characterization of $CO_2$ storage sites.

## 1. Introduction

The rising level of atmospheric $CO_2$ has gained widespread attention due to its perceived role in global warming. More than 190 countries have accepted the Paris agreement to maintain the increase in the average global temperature below 2°C above the levels during pre-industrial era (Gao *et al.* 2017). To meet this demand, the geological sequestration of $CO_2$ is an option available for these countries to explore, as there are sequestration projects already underway in different parts of the globe including Sleipner field in Norway (Furre *et al.* 2017) and Decatur, Illinois, US (Kaven *et al.* 2014). Due to the energy contributions from fossil fuel, $CO_2$ geological sequestration remains a viable mitigation strategy until there are technologies to significantly reduce or eliminate $CO_2$ emissions while the global energy demands are met. Among the geological storage options such as unminable coal seams and depleted oil and gas reservoirs, deep saline aquifers possess a large storage capacity and are well distributed across the globe (IPCC 2005).

$CO_2$ injected into a deep saline aquifer undergoes several processes. The injected gas exists in a supercritical form, when it is more dense than the gas phase $CO_2$ but less than the ambient formation brine, due to the temperature and pressure in a typical storage condition (Shariatipour *et al.* 2016). As a result, the supercritical $CO_2$ (sc$CO_2$) rises and spreads beneath the sealing caprock due to a combination of buoyancy and regional aquifer background flow. A sealing caprock with a high capillary entry pressure offers a structural trap that physically prevents the less dense and less viscous sc$CO_2$ from leaking into the overburden formations. At the trailing end of the migrating sc$CO_2$ plume, the formation brine snaps off and immobilizes sc$CO_2$ within the pore spaces during an imbibition process in a capillary trapping mechanism (Emami-Meybodi *et al.* 2015). The spreading plume and the capillary trapped sc$CO_2$ in contact with the formation brine eventually dissolve, leading to the downward migration of the brine saturated with $CO_2$ and subsequent geochemical reactions between the saturated solution and the rock minerals (IPCC 2005). These processes occur at different spatial and temporal scales, imposing several non-trivial modelling challenges when they are to be tracked simultaneously. This often result in the simplification of focusing on one or more of the processes at a time to understand the subsurface behaviour of $CO_2$. Our attention in this study is on $CO_2$ dissolution since brine saturated with dissolved $CO_2$, being heavier than the resident brine, sinks to the bottom of the aquifer, reducing the chance of a possible leakage of $scCO_2$ through possible pathways including faults, fractures or abandoned wells.

The dissolution of $CO_2$ in brine relies on the mechanisms of diffusion and convection to transport the solute towards the bottom of the aquifer. During the diffusive transport of the dissolved $CO_2$, a diffusive boundary layer of $CO_2$ rich brine, which is heavier than the underlying resident brine, is created near the top of the aquifer. The subsequent growth of the gravitationally unstable diffusive boundary layer can setup convective instabilities. The convection of the $CO_2$-saturated brine towards the bottom of the aquifer, initiated by the density driven instability, accelerates the rate of dissolution of $scCO_2$, which is faster than the rate due to diffusion alone, and improves $CO_2$ storage security (Hassanzadeh *at al. 2005*).

Studies on convective instability during the geological sequestration of $CO_2$ is preceded by the historical studies on thermal convection due to the apparent similarities in both problems. One of the earliest studies of natural convection in a pure fluid with a fixed temperature at the top and bottom boundaries of a horizontal layer are the experiments of Bernard and the theoretical analyses by Rayleigh (Rayleigh 1916), generally described as the Rayleigh-Bernard problem. Fluid flow in Rayleigh-Bernard problem is governed by the Navier-Stokes equations. This problem in fluids was extended by theoretical and experimental investigations to a fluid-saturated porous medium by Horton and Rogers (1945) and Lapwood (1948), commonly referred to as the Horton-Rogers-Lapwood problem. In a fluid-saturated porous medium, the hydrodynamic equation of fluid motion is described by Darcy's law. These earlier studies are based on the dependence of density on temperature but were later extended to the cases where density depends on salinity in the theoretical and experimental studies by Wooding (1959). Subsequently, numerous works outlined in a detailed overview on the modelling of density driven flows in porous media for the thermal and saline cases were reported (Holzbecher 1998). Although the injection of $CO_2$ for enhanced oil recovery provided some understanding on the geological sequestration of $CO_2$ (Orr and Taber 1984), one of the earliest studies of convection due to $CO_2$ solute in an aquifer column was presented by Lindeberg and Wessel-Berg (1997). They considered a homogeneous reservoir with a constant $CO_2$ concentration at the top and bottom boundaries similar to the constant temperature boundary conditions used in the Horton-Rogers-Lapwood problem (Lindeberg and Wessel-Berg 1997). However, they acknowledged that the steady density gradient assumed in their work is to provide a first insight since a zero $CO_2$ flux is more appropriate at the bottom boundary in the context of $CO_2$ sequestration, and that at the onset of convection, the background density gradient will still be evolving in time. Following the procedure in Horton and Rogers (1945) to obtain a physical insight, they found that the effect of the geothermal gradient on convection could be neglected in the presence of a concentration gradient when there are orders of magnitude difference between the thermal and molecular diffusivities.

Since this pioneering study (Lindeberg and Wessel-Berg 1997), $CO_2$ convective instability has gained improved attention – motivating studies with appropriate boundary conditions in homogeneous systems, based on average formation properties, and in heterogeneous systems, mostly with a variable permeability field, by theoretical investigations, numerical simulations, and experimental studies. In heterogenous media, a range of permeability structure exists, but there are relatively more studies that focus on permeability distributions that are either random or are in form of dispersed impermeable barriers (Farajzadeh *et al.* 2010, Lindeberg and Wessel-Berg 2011, Ranganathan *et al.* 2012, Elenius and Gasda 2013, Green and Ennis-King 2014, Taheri *et al.* 2018). Besides those geological configurations, many aquifers, being naturally heterogeneous, are composites of small and large patches of different permeabilities. A patch could form a continuous permeable layer since sedimentary rocks (including saline aquifer formations) exists naturally in layers of distinguishable depositional facies (*Campbell 1967*).

Inspired by the existence of layering in various physical settings, the onset of instability has been studied in layered porous media with a steady (McKibbin and O'Sullivan 1981, Rees and Storeletten 2019) and transient base profiles (Rapaka *et al.* 2009, Daniel *et al.* 2015, Ryoo and Kim 2018). Rapaka *et al.* (2009) employed a Galerkin technique to reduce linearized governing equations to a set of coupled ordinary differential equations. They used non-modal stability analysis for the growth of small perturbations for a sinusoidally varying log-permeability field. Daniel *et al.* (2015) investigated the onset of convective instability in a periodic and smoothly varying log-permeability field using initial value problem (IVP) for a single Rayleigh number (*Ra = 500*), the dimensionless number that describes the relative importance of diffusion and convection in the system. Periodic permeability profiles introduce multiple permeability oscillations thereby obscuring the individual effects of either a low-permeability region or permeability gradients. In Ryoo and Kim (2018), the stability of a fluid in a porous medium with a monotonically increasing permeability field varying exponentially in the vertical direction is analysed using linear stability analysis (LSA) and direct numerical simulation.

Despite these advancements, the following knowledge gaps remain on the study of the onset of transient convective instability in a variable permeability field: (i) the effect of *Ra* for a non-monotonic non-periodic permeability profile, (ii) the role of the location of the concentration perturbation peak, and (iii) the impact of a sharp permeability change.

In this paper, we use LSA with QSSA to characterize the influence of a variable permeability field, with smooth and sharp (layered) transitions – due to a low-permeability layer embedded in a homogeneous system, on the onset of convective instability. Formation permeability is proportional to flow velocity in Darcy's law which implies that permeability variation would have non-trivial effects on the convective dynamics. We investigate the stability of this non-monotonic, non-periodic variable permeability field using QSSA, justifying our method with the initial Value Problem (IVP) approach as well as the Dominant mode analysis (Daniel *et al.* 2015). For the first time, a scaling for the location of the peak of the concentration eigenfunction for homogeneous systems is derived, which provides insight on how a variable permeability profile affects the onset time. Our analysis for both smooth and abrupt permeability transitions reveal that the influence of the permeability variation on the onset time declines at low *Ra* and depends on the interaction between the permeability profile and concentration perturbations. We show that the onset time is a non-monotonic function of the location of the embedded low- permeability layer.

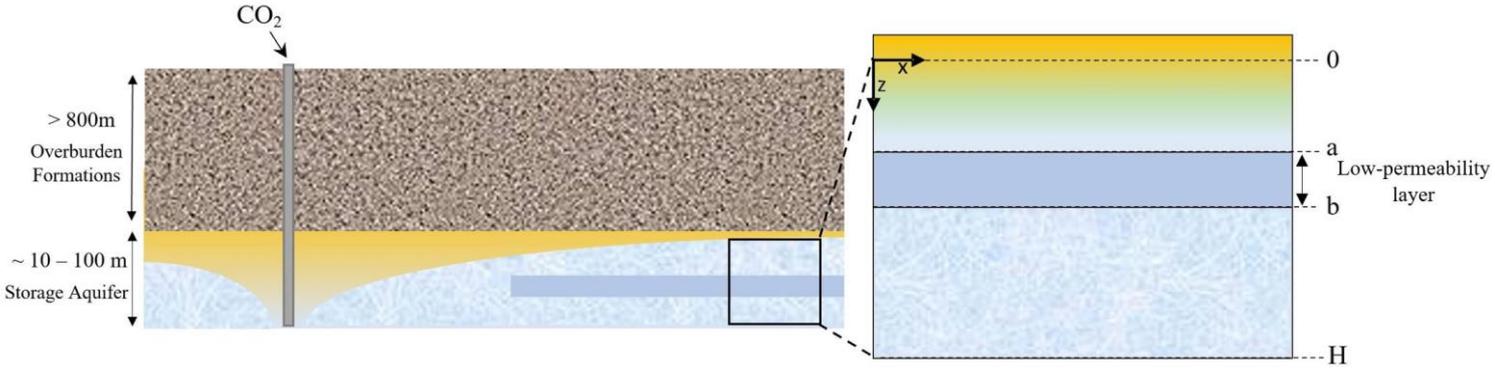

*Figure 1: A conceptual schematic of the CO$_2$ injection site under consideration. After injection, mobile scCO$_2$ spreads away from the wellbore within the permeable storage layer. Due to its smaller density, the scCO$_2$ rises and accumulates underneath the caprock, dissolving into the brine. The colors indicate CO$_2$ declining from yellow to blue.*

## 2. Model description and governing equations

We consider a horizontal, two-dimensional, isotropic, fluid-saturated porous medium with a vertically varying permeability. The medium is assumed to be laterally infinite. The top and bottom boundaries, both impermeable, have a constant source of dissolved CO$_2$ and no-flux, respectively. The fluid saturating the medium, transporting dissolved CO$_2$, is assumed to be in single-phase, incompressible, and quiescent. The medium is assumed to be isothermal, and rock-fluid geochemical reactions are ignored. The application and justification of these assumptions are outlined in the literature (Lindeberg and Wessel-Berg 1997, Ennis-King and Paterson 2005, Riaz *et al.* 2006, Xu *et al.* 2006, Hassanzadeh *et al.* 2007, Daniel *et al.* 2015, Raad and Hassanzadeh 2015, Ryoo and Kim 2018). This study focuses on *Ra ≤ 500* which is typical for aquifers within the range of the parameter values of permeability (≈ *10$^{-16}$ – 10$^{-13}$*

$m^2$), viscosity (≈ 0.5 – 0.8 mPas), porosity (≈ 0.06 – 0.18), diffusivity (≈ 2.6 – 7.6 × $10^{-9}$ $m^2/s$), thickness (≈ 4m – 81m), and density difference between the $CO_2$ saturated and the unsaturated aquifer fluid (≈ 1.51 – 6.15 $kg/m^3$) (*Hassanzandeh et al. 2007*), but note that it can be very large ($Ra \sim 10^5$) in other formations.

The governing equations for the single-phase flow of an incompressible fluid in a porous medium with a varying permeability field can be written as (Riaz *et. al.* 2006, Xu *et al.* 2006, Hassanzadeh *et al*. 2007):

$$\frac{\partial u}{\partial x} + \frac{\partial v}{\partial z} = 0 \qquad (1)$$

$$\frac{u}{k(z)} = -\frac{1}{\mu}\frac{\partial p}{\partial x} \qquad (2)$$

$$\frac{v}{k(z)} = -\frac{1}{\mu}\left(\frac{\partial p}{\partial z} - (1 + \beta\Delta c)\rho_0 g\right) \qquad (3)$$

$$\emptyset\frac{\partial c}{\partial t} + u\frac{\partial c}{\partial x} + v\frac{\partial c}{\partial z} = \emptyset D\left(\frac{\partial^2 c}{\partial x^2} + \frac{\partial^2 c}{\partial z^2}\right) \qquad (4)$$

where $u$ and $v$ are the Darcy velocity components in the horizontal and vertical directions respectively, $k(z)$ is the variable permeability, $p$ is the pressure, $c$ is the $CO_2$ dissolved concentration, $g$ is the gravity, μ is the viscosity, $\phi$ is the porosity, and $D$ is the effective diffusion coefficient of $CO_2$ in the porous medium saturated with brine, which captures the diffusivity of $CO_2$ in brine and the effect of the rock tortuosity. Under $CO_2$ storage condition, the permeability variations in geological formations are generally orders of magnitude larger than the variations in the fluid and other rock properties such as viscosity (Bando *et al. 2004*, Kumagai and Yokoyama 1999), effective diffusion coefficient (Li *et al.* 2021), and porosity (Reece *et al.* 2012, Casey *et al.* 2013). These parameters are therefore assumed to be constant, to highlight the effect of the variable permeability, as this study is restricted to a permeability variation within an order of magnitude when the corresponding change in other rock properties are very small.

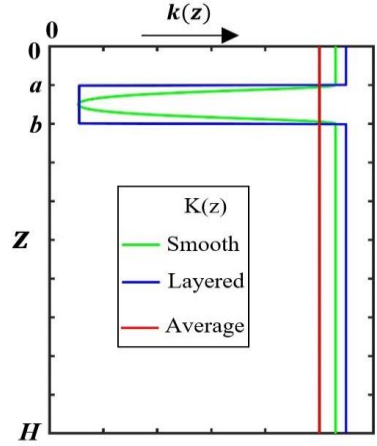

*Figure 2: The smooth and layered variable permeability field under consideration indicating the interval $a < z < b$ containing the embedded low-permeability layer. The permeability profiles are selected such that the variation is within an order of magnitude, and the average permeability for the smooth and layered permeability fields are equal.*

We start by assuming that the formation has a smooth, piecewise, differentiable permeability field $k(z)$ in the form:

$$\ln k(z) = \begin{cases} \ln k_H, & z < a \\ \ln(k_H/3.16) + \psi_0 \cos(\omega_k z + \beta), & a < z < b \\ \ln k_H, & b < z < H \end{cases}, \quad (Figure\ 2) \text{ where } k_H$$

represents the local permeability near the top and bottom boundaries. $\omega_k$ is the permeability wavenumber. The permeability phase, $\beta$, is selected to vary the interval $a < z < b$, denoting the low-permeable layer, and the permeability variation within the interval is such that the natural logarithm of the permeability profile varies sinusoidally (Ryoo and Kim 2018, Daniel *et al.* 2015, Sajjadi and Azaiez 2013, Rapaka *et al.* 2009, De Wit and Homsy 1997). By choosing $\psi_0 \approx 1.15$, we limit the permeability change within an order of magnitude, noting that the profile bears the essential features of a medium containing a low-permeability layer of a relatively small spatial extent (Campbell 1967). Boussinesq approximation is made so that the density dependence on the concentration is applied only to the buoyancy term

($\rho g$) in (3). Density is linearly dependent on the concentration in the expression $\rho = \rho_0(1 + \beta c)$, where $\rho_0$ is the original brine density, and $\beta$ is the coefficient of density variation with concentration.

The following dimensionless variables are introduced to the governing equations (1) – (4) with the hydrostatic term ($\rho_0 gz$) subtracted from the pressure: $\hat{x} = x/H$, $\hat{z} = z/H$, $\hat{a} = a/H$, $\hat{b} = b/H$, $\hat{c} = c/c_s$, $\hat{\omega}_k = \omega_k H$, $\hat{u} = Hu/\phi D$, $\hat{v} = Hv/\phi D$, $\hat{t} = Dt/H^2$, $\hat{p} = k_{avg}(p - \rho_0 gz)/\phi\mu D$, $\hat{k} = k(z)/k_{avg}$ to obtain the dimensionless governing equations:

$$\frac{\partial \hat{u}}{\partial \hat{x}} + \frac{\partial \hat{v}}{\partial \hat{z}} = 0 \quad (5)$$

$$\frac{\hat{u}}{\hat{k}} = -\frac{\partial \hat{p}}{\partial \hat{x}} \quad (6)$$

$$\frac{\hat{u}}{\hat{k}} = -\frac{\partial \hat{p}}{\partial \hat{z}} + Ra\hat{c} \quad (7)$$

$$\frac{\partial \hat{c}}{\partial \hat{t}} + \hat{u}\frac{\partial \hat{c}}{\partial \hat{x}} + \hat{v}\frac{\partial \hat{c}}{\partial \hat{z}} = \frac{\partial^2 \hat{c}}{\partial \hat{x}^2} + \frac{\partial^2 \hat{c}}{\partial \hat{z}^2} \quad (8)$$

where $c_s$ is the solute concentration of $CO_2$ in brine at the top boundary, $k_{avg} = H^{-1}\int_0^H k(z)$ and $H$ is the overall thickness of the domain. $Ra = \frac{k_{avg}\Delta\rho gH}{\phi\mu D}$ is the Rayleigh number describing the global system based on the average ($k_{avg}$) of the variable permeability while the other parameters have a constant value (Rapaka *et al.* 2009, Daniel *et al.* 2015). The dimensionless permeability field is

$$\ln \hat{k} = \begin{cases} 0.058, & 0 < \hat{z} \leq \hat{a} \\ -1.09 + \psi_0 \cos(\hat{\omega}_k \hat{z} + \beta), & \hat{a} < \hat{z} \leq \hat{b}, \\ 0.058, & \hat{b} < \hat{z} \leq 1 \end{cases}$$ when $\hat{\omega}_k = 20\pi$. The pressure term is eliminated by taking the $z$ – derivative of (6) and $x$ – derivative of (7), and applying equation (5) to the difference of the x and z derivative of the Darcy's equations to obtain

$$\frac{\partial^2 \hat{v}}{\partial \hat{x}^2} + \frac{\partial^2 \hat{v}}{\partial \hat{z}^2} - \frac{1}{\hat{k}} \frac{d\hat{k}}{d\hat{z}} \frac{\partial \hat{v}}{\partial \hat{z}} = \hat{k} Ra \frac{\partial^2 \hat{c}}{\partial \hat{x}^2} \qquad (9)$$

3. **Linear Stability Analysis (LSA)**

LSA is performed for a porous medium with a vertically varying permeability by decomposing the flow and transport variables $\hat{u}, \hat{v}, \hat{p}$ and $\hat{c}$ in the governing equations (5) – (8) into a transient base state and perturbation in the form $\hat{A}(x, z, t) = \hat{A}_b(z, t) + \hat{A}'(x, z, t)$. The subscript (b) and superscript (′) refer to the base state and the perturbed variable respectively; thus, $\hat{A}$ represents each component of the Darcy's velocity ($\hat{u}$ and $\hat{v}$), dissolved $CO_2$ concentration ($\hat{c}$) and pressure ($\hat{p}$). For the base state problem, we set $\hat{u}_b = \hat{v}_b = 0$ in the governing equations (5) – (8) and note that $\frac{\partial \hat{c}_b}{\partial \hat{x}} = 0$ since the base state concentration ($\hat{c}_b$) does not vary in the x-direction. The base concentration $\hat{c}_b$ satisfies the diffusion equation $\frac{\partial^2 \hat{c}_b}{\partial \hat{z}^2} = \frac{\partial \hat{c}_b}{\partial \hat{t}}$, the outer surface boundary conditions earlier described, and the initial condition $\hat{c} = 0$ in $0 < \hat{z} < 1$ for $\hat{t} = 0$. Its solution may be found using separation of variables as $\hat{c}_b(z, t) = 1 - \frac{4}{\pi} \sum_{m=1}^{\infty} \frac{1}{(2m-1)} e^{-\left(\frac{(2m-1)\pi}{2}\right)^2 \hat{t}} \sin \frac{(2m-1)\pi \hat{z}}{2}$. Substituting for the decomposition of the flow and transport variables into (8) and (9) and assuming that the magnitude of the perturbation variables is sufficiently small – so that the higher order perturbation terms can be ignored, the perturbation equations can be written as:

$$\frac{\partial \hat{c}'}{\partial \hat{t}} + \hat{v}' \frac{\partial \hat{c}_b}{\partial \hat{z}} = \frac{\partial^2 \hat{c}'}{\partial \hat{x}^2} + \frac{\partial^2 \hat{c}'}{\partial \hat{z}^2} \qquad (10)$$

$$\frac{\partial^2 \hat{v}'}{\partial^2 \hat{x}} + \frac{\partial^2 \hat{v}'}{\partial^2 \hat{z}} - \frac{1}{\hat{k}} \frac{d\hat{k}}{d\hat{z}} \frac{\partial \hat{v}'}{\partial \hat{z}} = \hat{k} Ra \frac{\partial^2 \hat{c}'}{\partial \hat{x}^2} \qquad (11)$$

Since the base state concentration profile in (10) is transient $\frac{\partial \hat{c}_b}{\partial \hat{z}}(\hat{z}, \hat{t})$, we employ the quasi-steady state approximation (QSSA), which is valid for the systems within the interval under consideration $Ra = 100 - 1000$ (Raad and Hassanzadeh 2016, Raad and Hassanzadeh 2015), to handle the time dependency of the base state by freezing the evolution of the base state concentration profile $\frac{\partial \hat{c}_b}{\partial \hat{z}}(\hat{z}, \hat{t}_0)$, at a given time $\hat{t}_0$. The perturbed variable is written in the form $(\hat{v}', \hat{c}')(\hat{x}, \hat{z}, \hat{t}) = (v^*, c^*)(\hat{z}, \hat{t}_0) e^{i\omega \hat{x} + \sigma(\hat{t}_0)\hat{t}}$ from Fourier decomposition where $\omega$ and $\sigma$ are the horizontal wavenumber and growth rate of the perturbations respectively. We substitute the expression for the perturbed variables into the equations (10) and (11) to obtain the following ordinary differential equations for the perturbations

$$v^* \frac{\partial \hat{c}_b}{\partial \hat{z}} = \left(\frac{d^2}{d\hat{z}^2} - \omega^2\right) c^* - \sigma c^* \qquad (12)$$

$$\left(\frac{d^2}{d\hat{z}^2} - \omega^2\right) v^* - \frac{1}{\hat{k}} \frac{d\hat{k}}{d\hat{z}} \frac{dv^*}{d\hat{z}} = -\omega^2 Ra \hat{k} c^* \qquad (13).$$

The associated boundary conditions are $v^* = c^* = 0$ at $\hat{z} = 0, \hat{t} > 0$ and $v^* = \frac{\partial c^*}{\partial \hat{z}} = 0$ at $\hat{z} = 1, \hat{t} > 0$. Equations (12) and (13) are variable coefficient differential equations, so we employ a finite difference discretization. Using finite difference, we pose the equations (12) and (13) in the matrix form:

$$\mathbf{diag}\left(\frac{\partial \hat{c}_b}{\partial z}\right) v^* = Ac^* - \sigma c^* \qquad (14)$$

$$(\boldsymbol{B} - \boldsymbol{G})v^* = -\omega^2 Ra \, \mathbf{diag}(\hat{k}) \, c^* \qquad (15)$$

where $c^*$ and $v^*$ are the concentration and velocity eigenfunctions respectively, $\boldsymbol{A}c^* = \left(\frac{d^2}{d\hat{z}^2} - \omega^2 * \boldsymbol{I}\right) c^*$ and $\boldsymbol{B}v^* = \left(\frac{d^2}{d\hat{z}^2} - \omega^2 * \boldsymbol{I}\right) v^*$, $\boldsymbol{G}v^* = \mathrm{diag}(\frac{1}{\hat{k}}\frac{d\hat{k}}{d\hat{z}})\frac{dv^*}{d\hat{z}}$ and $\boldsymbol{I}$ is an identity matrix. The notation diag(.) represents a diagonal matrix with the elements of the given vector, while $\frac{d}{d\hat{z}}$ $and$ $\frac{d^2}{d\hat{z}^2}$ are the matrices representing the first and second order central difference operators, respectively. The elements in the top and bottom rows of the coefficient matrices $A$ and $B$ represents the Dirichlet boundary condition on concentration and Neumann condition on velocity, respectively.

Rearranging, we have the eigenvalue problem:

$$\sigma c^* = \left(A - (-\omega^2)Ra\, \mathbf{diag}(\hat{k})(B-G)^{-1}\mathbf{diag}\left(\frac{\partial \hat{c}_b}{\partial z}\right)\right)c^* \qquad (16).$$

The growth rate for different wavenumbers is obtained by solving the eigenvalue problem (16). For the homogeneous case, all the elements of the vector $\hat{k}(z) = 1$, and $G = 0$ leading to a simplified eigenvalue problem $\sigma c^* = \left(A - (-\omega^2)Ra B^{-1}\frac{\partial c_b}{\partial z}\right)c^*$, which was derived in Raad and Hassanzadeh (2015).

The perturbation growth rate is measured by the maximum eigenvalue of the coefficient matrix $(A-\omega^2)Ra\, \mathrm{diag}(\hat{k})(B-G)^{-1}\mathrm{diag}\left(\frac{\partial \hat{c}_b}{\partial \hat{z}}\right)$, and the maximum value indicates a stable or unstable system when it is negative or positive, respectively (Raad and Hassanzadeh 2015). Our LSA numerical computation for the variable permeability field ($\hat{k}(z) = e^{\left(-\frac{\delta d}{Ra}\hat{z}\right)}$) with $\frac{\delta d}{Ra} = -0.01$, $\omega = 0.07$, $\hat{t} = 500$ gives a growth rate of *0.0075*, which is in close agreement with the results reported by Ryoo and Kim (2018). Our computation qualitatively agrees with the results using the IVP and the dominant mode (Daniel et al. 2015) methods in which increasing $\sigma^2$ increases stability for $\gamma = 0$ but decreases stability for $\gamma = \pi$ due to the magnitude of the local permeability around the neighbourhood of the peak of the concentration eigenfunction (Figure 3). This agreement is found for both (i) $\sigma^2 = 0, 0.1$, and 1.0 when $\gamma = 0$ and (ii) $\sigma^2 = 0, 0.35$ and 1.0 when $\gamma = \pi$, with $\hat{k}(z) = \sqrt{2\sigma^2}e^{(\cos(m\hat{z}+\gamma))}$ and $m = 6\pi$. The agreement of our computations with LSA results (Ryoo and Kim 2018) and IVP results (Daniel *et al.* 2015) for a variable permeability field in a porous medium validates our LSA numerical algorithm.

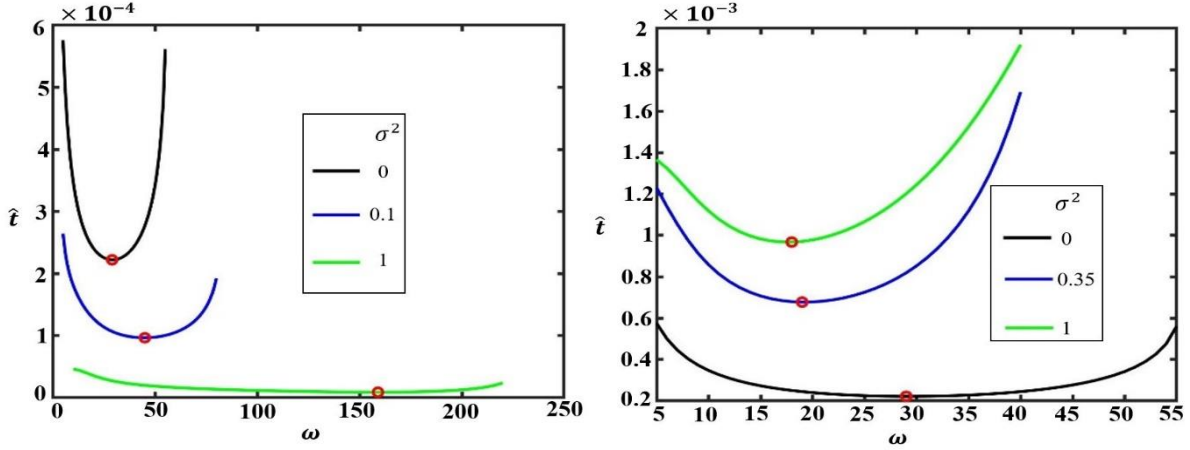

*Figure 3: The stability behaviour of the variable permeability field ( $\hat{k}(z) = \sqrt{2\sigma^2}e^{(\cos(m\hat{z}+\gamma))}$ ) where $m = 6\pi$ for various values of $\sigma^2$ for a) $\gamma = 0$ and b) $\gamma = \pi$.*

**3.1 Layered Permeability Profile Approximation**

We now show how the finite difference method extends the validity of this formulation to a discontinuous permeability field when the layered permeability profile (*Figure 1*) is approximated by a non-arbitrary continuous function $(\hat{k})$. We introduce the condition based on the pressure continuity across the domain: $\frac{1}{\hat{k}}\frac{\partial \hat{v}}{\partial \hat{z}} = \frac{\partial^2 \hat{p}}{\partial^2 \hat{x}}$, derived by taking the *x*-derivative of (6) and applying mass conservation, to obtain the condition in terms of the combination of the permeability and vertical velocity. The permeability of the top and bottom layers is denoted as $k_H$, and $k_L$ is the permeability of the low-permeable layer. In dimensionless form, the elements in the vector $\hat{k}$ includes $\hat{k}_H = \frac{k_H}{k_{avg}}$, and $\hat{k}_L = \frac{k_L}{k_{avg}}$. For a step change in the permeability across an $n^{th}$ internal boundary separating a medium *n* from *n+1*, the condition on pressure at the boundary requires that $\frac{1}{\hat{k}_n}\frac{\partial \hat{v}_n}{\partial \hat{z}} = \frac{1}{\hat{k}_{n+1}}\frac{\partial \hat{v}_{n+1}}{\partial \hat{z}}$ since the pressure is continuous across the internal boundaries. In a finite difference framework, we can implement the internal boundary condition on pressure by choosing specific values in the vectors $\hat{k}$ and $\frac{d\hat{k}}{d\hat{z}}$ around the internal boundaries. Assuming $\hat{z}_n$ is a node in the discretization of $\hat{z}$ near $\hat{z}_n = \hat{a}$ having neighbours $\hat{z}_{n-1}$ and $\hat{z}_{n+1}$, the finite difference approximation of the linearized equation (*13*) at this node as $\delta\hat{z} \to 0$ is given by: $\frac{v_{n-1}^* - 2v_n^* + v_{n+1}^*}{\delta\hat{z}^2} - \omega^2 v_n^* - \frac{1}{\hat{k}}\frac{d\hat{k}}{d\hat{z}}\frac{v_{n+1}^* - v_{n-1}^*}{2\delta z'} = -\omega^2 Ra\hat{k}c_n^*$. By using the expression $\hat{k} = \frac{\hat{k}_n + \hat{k}_{n+1}}{2}$ and $\frac{d\hat{k}}{d\hat{z}} = \frac{\hat{k}_{n+1} - \hat{k}_n}{\delta\hat{z}}$, the finite difference approximation at the boundary node gives the first order approximation to the boundary condition on pressure at the internal boundaries $\frac{1}{\hat{k}_n}\frac{v_n^* - v_{n-1}^*}{\delta\hat{z}} = \frac{1}{\hat{k}_{n+1}}\frac{v_{n+1}^* - v_n^*}{\delta\hat{z}} + O(\delta\hat{z})$.

## 4 Results and Discussions

We analyse the effect of an embedded low permeable layer on the onset of transient convective instability in a fluid saturated porous medium based on the smooth, layered and the average permeability field (Figure 2).

### 4.1 Homogeneous Systems ( $\hat{k}_{avg}$ )

During the early time, the growth rate is negative $(\sigma < 0)$, which represents a decay in the evolution of the introduced perturbations indicating a stable system. However, at larger times, the growth rate becomes positive, which first happens at the critical onset time and the critical wavenumber (*Figure 4a*). The relationship $\hat{t}_o = \alpha_0 Ra^{-2}$ between the critical time of the onset of convective instabilities, which we simply refer to as the onset time, and the Rayleigh number for homogeneous systems is obtained for *Ra = 100 – 1000*, with a prefactor $\alpha_0 \approx$ 56.

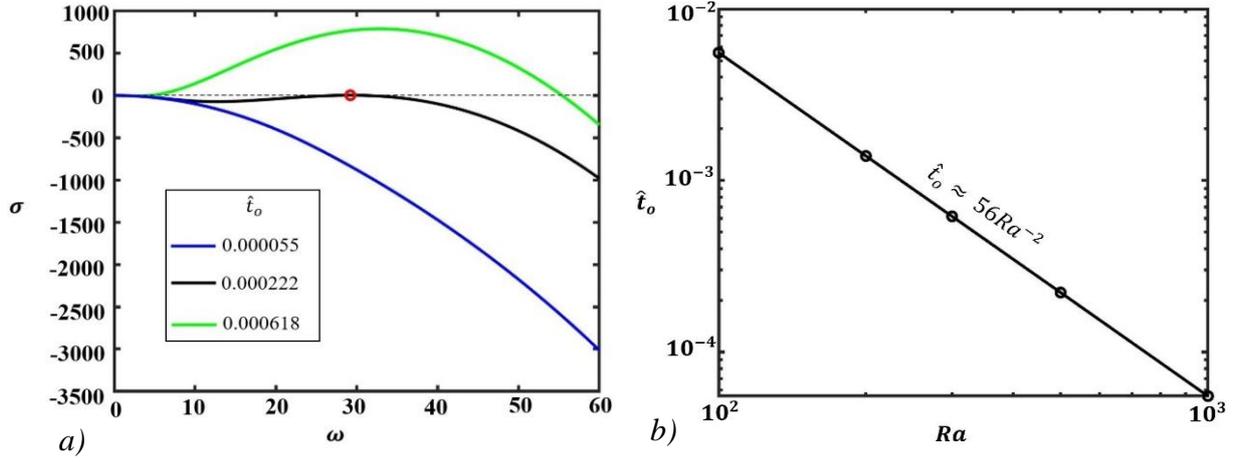

*Figure 4. (a) The perturbation growth rate versus wavenumber at various times showing the critical onset time and wavenumber in the red circle for Ra =500 (b) The relationship between the critical onset time of convective instability and Rayleigh number for homogeneous systems.*

The obtained scaling is in close agreement with previous methods including LSA QSSA ($\alpha_0 \approx 56$; Raad and Hassanzadeh 2015), LSA IVP ($\alpha_0 \approx 75$; Xu et al. 2006), LSA dominant mode method ($\alpha_0 \approx 146$; Riaz et al. 2006) and numerical simulations ($\alpha_0 \approx 160$; Elenius and Johannsen 2012). The prefactor ($\alpha_0$) depends on the method of analysis. Recalling that $Ra = \frac{k_{avg}\Delta\rho g H}{\phi \mu D}$ and $\hat{t} = \frac{Dt}{H^2}$ where $\Delta\rho = \rho_0 \beta C_s$, the dimensional critical time is $t_o = \alpha_0 \left(\frac{\phi \mu \sqrt{D}}{k_{avg}\Delta\rho g}\right)^2$.

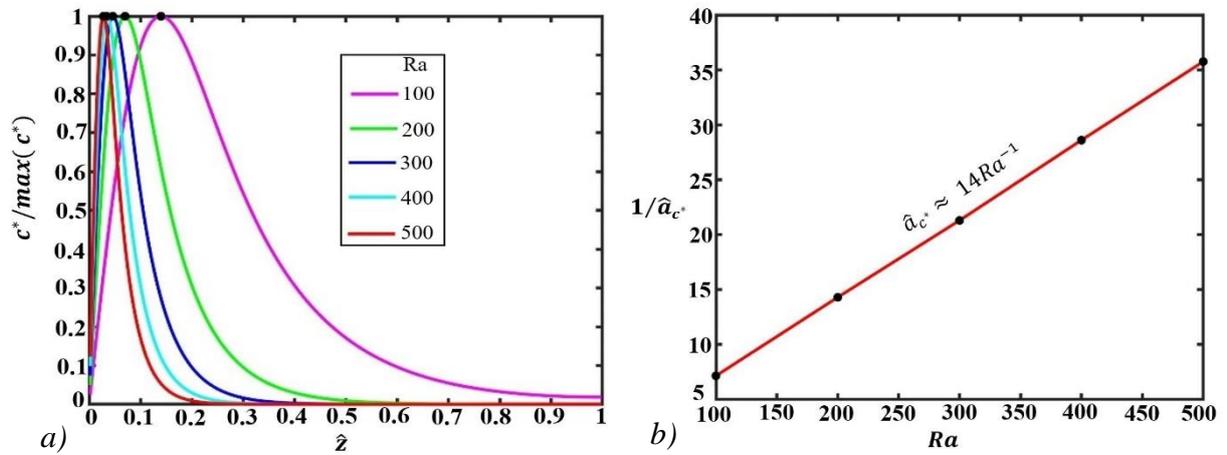

*Figure 5: a) The peak of the concentration eigenfunction which is a reference for the spatial location of where the instabilities are formed at the critical onset time. b) The scaling relationship between the location of the peak and Ra.*

The concentration eigenfunction describes the perturbation profile (*Daniel et al. 2015*), and the location of its peak ($\hat{a}_{c^*}$) is a reference point for the spatial location of the onset of the convective instability. The location of the peak scales as $\hat{a}_{c^*} = \gamma_0 Ra^{-1}$, where $\gamma_0 \approx 14$, implying that instabilities are formed in large *Ra* systems near the top boundary. These observations (Figure 4 and Figure 5) support the previous claim which specifies that small disturbances are not readily propagated in low permeability systems (Lindeberg and Wesselberg 2011, Cheng *et al.* 2012).

## 4.2 Variable permeability system ($\hat{k}(\hat{z})$)

We describe the behaviour of the onset time in a medium with a variable permeability field due to the existence of a low-permeability layer, comparing the role of a smooth permeability transition to that of a layered case. In addition, the dependence of the onset time on *Ra* and the location $\hat{a}$ of the low-permeability layer is explained. This analysis is restricted to the cases in which the embedded low-permeability layer has a constant dimensionless thickness ($\hat{b} - \hat{a} = 0.1$), and the permeability variations are only within an order of magnitude. The location of the top of the low-permeability layer, â, is varied to investigate the effect of the resulting permeability field on the onset of convective instability. An increasing value of â represents shifting the low permeability layer towards the bottom of the domain.

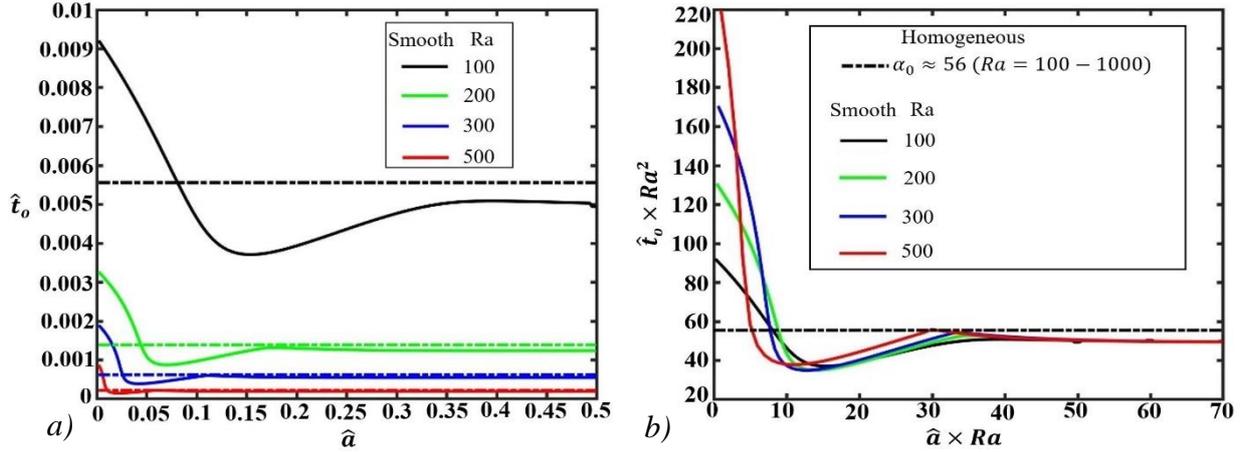

*Figure 6: The relationship between the onset time and the distance of the embedded low-permeability layer from the top boundary (â) for Ra = 100 – 500 when the length scale is a) H b) H/Ra for the smooth case. The homogeneous case is shown by the dashed lines.*

The role of the embedded low-permeability layer on the onset time is twofold: a) it introduces two regions with the sudden permeability transition from $\hat{k}_H$ to $\hat{k}_L$ around the top (â) and bottom ($\hat{b}$) of the layer that contributes certain disturbances to the perturbations that triggers instabilities, and b) it impedes flow and the formation of instabilities due to its local, low permeability $\hat{k}_L$ since the flow velocities scales with permeability, $[\hat{u}, \hat{v}] \sim \hat{k}$, from Darcy's law. The first role of contributing disturbances gains relevance when the location of the peak of the concentration eigenfunction ($\hat{a}_{c^*}$) approaches the region where there is an abrupt permeability change either from the top or bottom ($\hat{a}_{c^*} \to \hat{a}$ or $\hat{b}$). The second role is significant when the formation of instabilities occurs below or well-within the low-permeability layer ($\hat{a}_{c^*} > \hat{a}$), after enough $CO_2$ has gone through the low-permeability layer purely by diffusion, which is a transport mechanism that does not depend on the layer permeability. In this scenario, the formation of instabilities depends on the effect of the layer's low permeability values on the flow or propagation of disturbances, and this dependence increases with *Ra* (Figure 6). However, when $CO_2$ is concentrated near the top boundary of the low-permeability layer ($\hat{a}_{c^*} \approx \hat{a}$), the effect of the concentration perturbations within or below the low-permeability is less important, making the effect of the disturbance contribution from the permeability change dominate as $\hat{a}_{c^*} \to \hat{a}$. Furthermore, when the low-permeability layer is far from the peak of the concentration eigenfunction ($\hat{a}_{c^*} \ll \hat{a}$), the onset time is similar to that of a system that has a homogeneous permeability field equivalent to that of the local top or bottom boundary ($\hat{k}_H$).

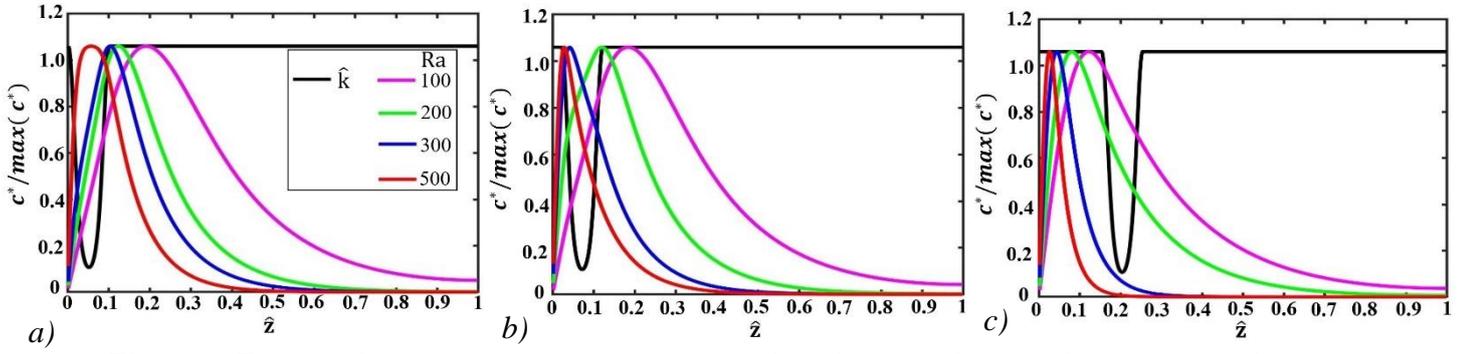

*Figure 7: The location of the peak of the concentration eigenfunction relative to the location of the top of the low-permeability layer for a) $\hat{a} = 0$, b) $\hat{a} = 0.022$ and c) $\hat{a} = 0.15$ for the smooth variation case.*

For $Ra = 100$, for instance, the peak of the concentration eigenfunction is below the low permeability layer ($\hat{a}$ or $\hat{b} < \hat{a}_{c^*}$) when the layer is near the top boundary (*Figure 7a*), and the onset time is due to a combination of the effect of the low-permeability layer on the formation of instabilities and the additional disturbances around the sudden permeability change. As the bottom of the low-permeability layer approaches the peak ($\hat{b} \to \hat{a}_{c^*}$) for large values of $\hat{a}$, the onset time is relatively shorter because the influence of the disturbance contribution from $\hat{b}$, where an abrupt permeability change occurs, is more significant (Figure 7b). This observation for $Ra = 100$ is consistent for the range of $Ra$ investigated in this study (Figure 6), but the onset time is more sensitive to the permeability variation at large $Ra$ (Figure 6b). Relative to the homogeneous system based on $k_{avg}$, the onset time is delayed since the formation of instabilities – when it occurs within or below the low-permeability layer – still depends on the propagation of disturbances within the low-permeability layer. On the other hand, when the instabilities are formed near the top boundary, *for example* in *Ra = 100*, the onset time is early compared to the homogeneous system based on $k_{avg}$ since the effect of the disturbances originating from $\hat{a}$ is the most important (Figure 7c). At this $\hat{a}$ (= *0.15*), the onset time for *Ra = 500* is equivalent to that with homogeneous permeability field of $\hat{k}_H$ since $\hat{a}_{c^*} \ll \hat{a}$ (Figure 7c).

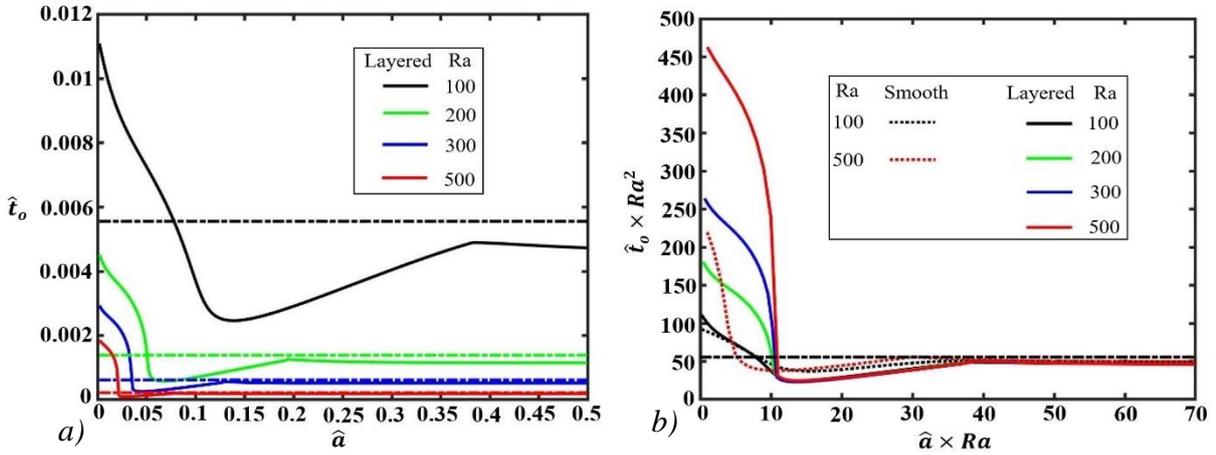

*Figure 8: The relationship between the critical onset time and the distance of the low-permeability layer from the top boundary ($\hat{a}$) for Ra = 100 – 500 when the length scale is a) H b) H/Ra for the layered case. The homogeneous case is shown by the dashed lines.*

As the permeability variation approaches sharp layering, the onset time shows a stronger dependence on the location of the low-permeability layer in comparison to the smooth permeability variation. The behaviour of the onset time is similarly due to the location of the peak of the concentration eigenfunction ($\hat{a}_{c^*}$) with respect to the location of the low-permeability layer ($\hat{a}$) (Figure 8 and Figure 9). This greater dependence is evident in the large prefactor range for the systems with layered permeability (Figure 8b). The location of the low-permeability layer that results in the earliest onset time scales as $\hat{a}Ra \approx 14$ which is similar to the relationship $\hat{a}_{c^*} \approx 14Ra^{-1}$ between location of the homogeneous peak of the concentration eigenfunction ($\hat{a}_{c^*}$) and $Ra$. This result shows that, for the strict permeability layering, the critical onset time is accelerated by ~2.3 ($\hat{t}_o \approx 24Ra^{-2}$) times earlier than homogeneous onset time when the location of the low-permeability layer coincides with the location of the homogeneous peak of the concentration eigenfunction $\hat{a} \approx \hat{a}_{c^*}$.

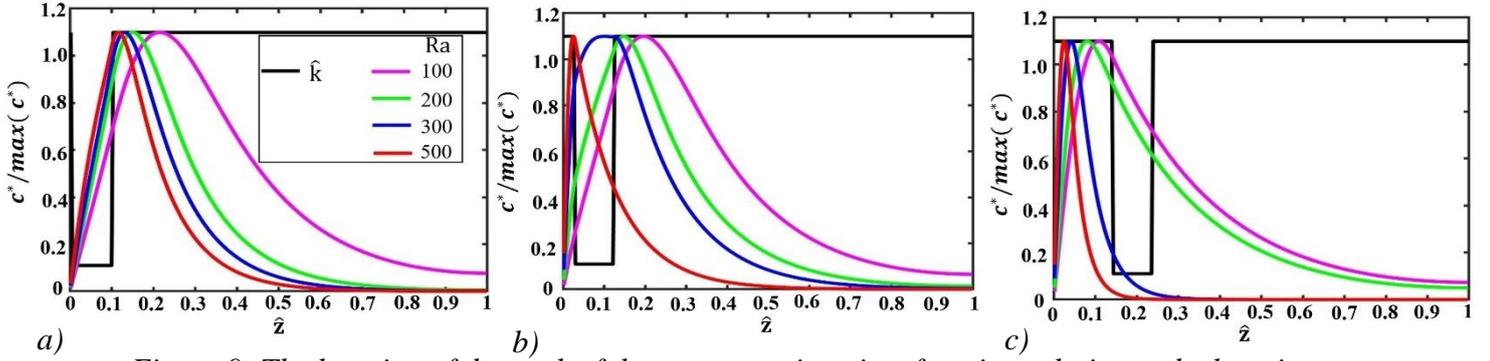

*Figure 9: The location of the peak of the concentration eigenfunction relative to the location of the top of the low-permeability layer for a) $\hat{a} = 0$, b) $\hat{a} = 0.028$ and c) $\hat{a} = 0.14$.*

### 4.3 Practical Implications

The potential implications of our findings are illustrated by computing the onset time using the following average formation properties: $H = 20m$, $k_H = 10^{-14}m^2$, $\mu = 0.00051\ Pas$, $D = 10^{-9}m^2/s$, $\Delta\rho = 5.7\ kg/m^3$, $\emptyset = 16\%$, and $g = 9.8\ m/s^2$, motivated by the low-permeability Krechba formation, In Salah gas field in Algeria (Elenius and Johanssen 2012), which has a Rayleigh number ($\approx 137$) within our range of study. This site is characterized by the stratification of different geologic layers, with complex reservoir heterogeneity influenced by fractures and faults (Durucan *et al.* 2011), which are not accounted for in this work. For the average homogeneous reservoir properties, the onset time is $38\ yr$. Using the idealized, sharp permeability layering comprising a 2m thick ($b - a = 0.1H$) low-permeability layer ($k_L = 10^{-15}m^2$) sandwiched within the homogeneous system at a depth of 2m ($a = 0.1H$), which coincides with the homogeneous concentration eigenfunction peak ≈ 2m ($a_{c^*} \approx 14Ra^{-1}H$), the onset time is ≈ 16 $yr$, due to $CO_2$ still concentrated at the top of the formation, getting near but has not gone through the low-permeability layer since $a \approx a_{c^*}$. However, when the low-permeability layer is placed near the caprock ($a = 0$), the onset time is delayed to ≈ 92 $yr$ as ($a_{c^*} > a$). Since the sharp permeability variation defines the limits, the onset time due to any similar smooth permeability variation within an order of magnitude difference will occur within the interval $16\ yr < t_o < 92\ yr$. These results make potential sites for $CO_2$ storage with low-permeability layers near the caprock less attractive, as they could delay the formation of convective instabilities.

## 5. Conclusion

We have presented a theoretical analysis of the effect of layering on the onset of transient convective instability in a porous medium with a variable, non-monotonic permeability field due to the existence of a low-permeability layer in a $CO_2$ storage reservoir. Linear stability analysis based on the quasi-steady state approximation (QSSA) is employed to handle the time-dependency of the diffusive base state. We show that the onset time is delayed when the peak of the concentration perturbation occurs below the low-permeability layer; in contrast, the onset happens earlier when the perturbation peak is near the top of the low-permeability layer. This effect is more evident in layered systems in which the permeability changes abruptly and is dependent on *Ra,* especially when the low-permeability layer is near the top. The substantial impact of a thin low-permeability layer on convective instability highlights the importance of detailed knowledge of the geological structure for identifying potential $CO_2$ storage sites.

## Acknowledgement

The authors gratefully acknowledge Petroleum Technology Development Fund (PTDF) in Nigeria for funding this study. The authors appreciate the insightful discussions and comments from Dr. Hassan Hassanzadeh and Nasser Sabet.